\documentclass[12pt]{iopart}
\usepackage{graphicx}
\expandafter\let\csname equation*\endcsname\relax
\expandafter\let\csname endequation*\endcsname\relax
\usepackage{amsmath}
\usepackage{mathabx}
\usepackage{cite}
\usepackage{lscape}
\usepackage{varwidth}
\usepackage{textcomp}
\usepackage{xcolor,soul}
\usepackage{color,soul}
\usepackage{algorithm}
\usepackage{algpseudocode}
\usepackage{bm}
\usepackage{esvect}
\usepackage{accents}
\usepackage{caption}

\newcommand*{\dt}[1]{%
  \accentset{\mbox{\large\bfseries .}}{#1}}

\begin{document}

\title[MICROSCOPE mission analysis]{MICROSCOPE mission analysis, requirements and expected performance}


\author{Pierre Touboul$^1$, Manuel Rodrigues$^1$, Gilles M\'etris$^2$, Ratana Chhun$^1$, Alain Robert$^3$, Quentin Baghi$^2$ \footnote{Current address: NASA Goddard Space Flight Center, Greenbelt, MD 20771, USA}, Emilie Hardy$^1$, Joel Berg\'e$^1$, Damien Boulanger$^1$, Bruno Christophe$^1$, Valerio Cipolla$^3$, Bernard Foulon$^1$, Pierre-Yves Guidotti$^3$ \footnote{Current address: AIRBUS Defence and Space, F-31402 Toulouse, France},  Phuong-Anh Huynh$^1$, Vincent Lebat$^1$, Fran\c{c}oise Liorzou$^1$, Benjamin Pouilloux$^3$ \footnote{Current address: KINEIS, F-31520 Ramonville Saint-Agne, France}, Pascal Prieur$^3$, Serge Reynaud$^{4}$}

\address{$^1$ DPHY, ONERA, Universit\'e Paris Saclay, F-92322 Ch\^atillon, France}
\address{$^2$ Universit\'e C\^ote d{'}Azur, Observatoire de la C\^ote d'Azur, CNRS, IRD, G\'eoazur, 250 avenue Albert Einstein, F-06560 Valbonne, France}
\address{$^3$ CNES, 18 avenue E Belin, F-31401 Toulouse, France}
\address{$^{4}$ Laboratoire Kastler Brossel, UPMC-Sorbonne Universit\'e, CNRS, ENS-PSL University, Coll\`ege de France, 75252 Paris, France}

\ead{pierre.touboul@onera.fr, gilles.metris@oca.eu, manuel.rodrigues@onera.fr}

\vspace{10pt}
\begin{indented}
\item[] Dec. 2020
\end{indented}

\begin{abstract}
The MICROSCOPE mission aimed to test the Weak Equivalence Principle (WEP) to a precision of $10^{-15}$. The WEP states that two bodies fall at the same rate on a gravitational field independently of their mass or composition. In MICROSCOPE, two masses of different compositions (titanium and platinum alloys) are placed on a quasi-circular trajectory around the Earth. They are the test-masses of a double accelerometer. The measurement of their accelerations is used to extract a potential WEP violation that would occur at a frequency defined by the motion and attitude of the satellite around the Earth. 
This paper details the major drivers of the mission leading to the specification of the major subsystems (satellite, ground segment, instrument, orbit...). Building upon the measurement equation, we derive the objective of the test in statistical and systematic error allocation and provide the mission's expected error budget.

\end{abstract}

\noindent{\it Keywords}: General Relativity, Experimental Gravitation, Equivalence Principle, Space accelerometers, Microsatellite.
%

\submitto{\CQG}
%
%
%

\section{Introduction}
The MICROSCOPE (French acronym: Micro-Satellite \`a tra\^in\'ee Compens\'ee pour l'Observation du Principe d'Equivalence) mission was defined in 2000 by the Centre National d'Etudes Spatiales (CNES), the Observatoire de la C\^ote d{'}Azur (OCA) and the Office National d'Etudes et de Recherches A\'erospatiales (ONERA). 
Designed to test the WEP in space, the satellite was launched into a low-Earth sun-synchronous orbit from Kourou on April 25, 2016 at an altitude of 710 km, and delivered science data for more than two years. 
In Ref. \cite{touboul17}, 7\% of the available data were used to provide first, intermediate results. 
No evidence for a violation of the WEP was found at $1.3\times{}10^{-14}$, one order of magnitude higher than MICROSCOPE's full mission target accuracy of $10^{-15}$ on the E\"otv\"os parameter.

The MICROSCOPE satellite was designed to provide an environment as stable as possible. It is finely controlled along its six degrees of freedom thanks to the Drag-Free and Attitude Control System (DFACS), detailed in Ref. \cite{robertcqg3}. The DFACS allows for several modes of operation: inertial pointing or spin mode. In spin mode, the satellite rotates about the instrument's $Y$-axis in the direction opposite to the orbital motion in order to increase the apparent rate of the Earth gravity field variation. With a frequency rate of rotation $f_{\rm{spin}_i}$, the EP frequency $f_{\rm{EP}_i}=f_{\rm{orb}}+f_{\rm{spin}_i}$, where $f_{\rm{orb}}$ is the satellite's orbital frequency.  Table \ref{tab_freq} lists the available frequencies (see Ref. \cite{rodriguescqg4} for details).

\begin{table}
\caption{\label{tab_freq} Main frequencies of interest.}
\begin{indented}
\item[]\begin{tabular}{@{}llll}
\br
Label & Frequency & Comment \\
\mr
$f_{\rm{orb}}$ & $1.6818\times{}10^{-4}$ Hz & Mean orbital frequency \\
\mr
$f_{\rm{spin}_2}$ & $\frac{9}{2}f_{\rm{orb}}$=$0.75681 \times{}10^{-3}$ Hz& Spin rate frequency 2 (V2 mode) \\
\mr
$f_{\rm{spin}_3}$ & $\frac{35}{2} f_{\rm{orb}}$=$2.94315 \times{}10^{-3}$ Hz& Spin rate frequency 3 (V3 mode) \\
\mr
$f_{\rm{EP}_2}$ & $0.92499 \times{}10^{-3}$ Hz& EP frequency in V2 mode  \\
\mr
$f_{\rm{EP}_3}$ & $3.11133 \times{}10^{-3}$ Hz& EP frequency in V3 mode \\
\mr
$f_{cal}$ & $1.22848 \times{}10^{-3}$ Hz& Calibration frequency \\
\br
\end{tabular}
\end{indented}
\end{table}

The satellite carries the Twin Space Accelerometers for Gravitation Experiment (T-SAGE) payload.
T-SAGE is composed of two sensor units called SUREF and SUEP. Each sensor unit is a double accelerometer whose test masses are two co-axial hollow cylinders. SUREF's test-masses are made of the same material (PtRh10) while SUEP's test-masses are made of different material (PtRh10 for the inner mass, Ti alloys for the outer mass). The principle of the accelerometers and the description of the instrument are detailed in Ref. \cite{rodriguescqg1}.
Each Sensor Unit (SU) is associated to a Front End Electronic Unit (FEEU) which contains the pick-up measurement, the voltage references, the electrode voltages and the whole instrument temperature acquisition system.

This paper details the per-flight error budget used to establish the requirement tree for all subsystems of the mission. Based on the measurement equation (Sect. \ref{sect_exp}), it is detailed in Sect. \ref{sect_req}.
After the flight, the requirements were verified by direct or indirect measurements or by analysis. An update of the error budget evaluated with flight inputs is presented at the end of the paper.


\section{Measurement equation} \label{sect_exp}

The acceleration of each test-mass can be given in the satellite moving frame by: 
\begin{equation} \label{eq_acc}
\frac{{\rm d}^2\vv{O_\Earth O_j}}{{\rm d}t^2}=\frac{{\rm d}^2\vv{O_\Earth O_{\rm sat}}}{{\rm d}t^2}+\vv{\dot \Omega} \times \vv{O_{\rm sat}O_j}+\vv{\Omega}\times \left(\vv{\Omega} \times \vv{O_{\rm sat}O_j} \right)+ 2 \vv{\Omega} \times  \frac{{\rm d}\vv{O_{\rm sat}O_j}}{{\rm d}t} + \frac{{\rm d}^2\vv{O_{\rm sat}O_j}}{{\rm d}t^2},
\end{equation}
with $O_\Earth$ the centre of the Earth as the centre of the Galilean frame, $O_{\rm sat}$ the satellite centre of gravity and $O_j$ the centre of the $j$th test-mass. $\vv{\dot \Omega}$ represents the angular acceleration of the satellite and $\vv{\Omega}$ its angular velocity. When the test-masses are servo-controlled, their relative motion to the satellite structure can be considered as null and thus the terms $\frac{{\rm d}\vv{O_{\rm sat}O_j}}{{\rm d}t}$ and $\frac{{\rm d}^2\vv{O_{\rm sat}O_j}}{{\rm d}t^2}$ can be neglected in the measurement bandwidth. However, they are considered when the test-mass position is biased with a sine signal for calibration \cite{hardycqg6, bergecqg7}.

In addition Newton's second law gives respectively the satellite and the test-mass acceleration as:
\begin{equation} \label{eq_sat}
\frac{{\rm d}^2\vv{O_\Earth O_{\rm sat}}}{{\rm d}t^2}=  \frac{\vv{F}_{\rm ext}}{M_{{\rm I_{sat}}}}+ \frac{\vv{F}_{\rm th}}{M_{{\rm I_{sat}}}}+\frac{M_{{\rm G_{sat}}}}{M_{{\rm I_{sat}}}} \vv{g}(O_{\rm sat}),
\end{equation}
\begin{equation} \label{eq_tm}
\frac{d^2\vv{O_\Earth O_j}}{dt^2}=  \frac{\vv{F}_{{\rm el}, j}}{m_{\rm I}}+ \frac{\vv{F}_{{\rm dis}, j}}{m_{\rm I}}+\frac{m_{{\rm G}_j}}{m_{{\rm I}_j}} \vv{g}(O_{j}),
\end{equation}
where $M_{{\rm G_{sat}}}$ and $m_{{\rm G}_j}$ are the gravitational masses of the satellite and of the $j^{th}$ test-mass, $M_{I_{\rm sat}}$ and $m_{{\rm I}_j}$ are the inertial masses of the satellite and of the test-mass, $\vv{g}(O_{\rm sat})$ (resp. $\vv{g}(O_j)$) is the Earth gravity acceleration at the centre of mass of the satellite (resp. test mass). The satellite undergoes non gravitational forces $\vv{F}_{\rm ext}$ such as atmospheric drag and Solar radiation pressure and thruster forces $\vv{F}_{\rm th}$. The test mass undergoes electrostatic forces $\vv{F}_{{\rm el}, j}$ and internal disturbing forces $\vv{F}_{{\rm dis}, j}$ such as those induced by the electrostatic stiffness, the gold wire stiffness, radiometer effect, ...\cite{chhuncqg5}. The acceleration measurement is inferred from the measurement of the voltages applied on the electrodes and from the estimated scale factor \cite{liorzoucqg2}. By combining Eq. (\ref{eq_acc}), Eq. (\ref{eq_sat}) and Eq. (\ref{eq_tm}), the resulting electrostatic acceleration $\vv{\gamma}_j$ of the $j$th mass is defined as the electrostatic force divided by the mass : 

\begin{equation} \label{eq_meas0}
\vv{\gamma}_j = \frac{\vv{F}_{{\rm el}, j}} {m_{{\rm I}_j}} = \vv{\gamma}_{\Earth, j} + \vv{\gamma}_{{\rm kin}, j} - \frac{\vv{F}_{{\rm dis}, j}}{m_{{\rm I}_j}} + \frac{\vv{F}_{\rm ext}}{M_{\rm I_sat}}+ \frac{\vv{F}_{\rm th}} {M_{\rm I_sat}},
\end{equation}
with the contribution to  the Earth's gravitational acceleration given by:
\begin{equation} \label{eq_gravacck}
\vv\gamma_{\Earth, j} = \frac{M_{{\rm G_{sat}}}}{M_{{\rm I_{sat}}}} \vv{g}(O_{\rm sat}) - \frac{m_{{\rm G}_j}}{m_{{\rm I}_j}} \vv{g}(O_j),
\end{equation}

where $\vv{\gamma}_{{\rm kin}, j}=\vv{\dot \Omega}\times \vv{O_{\rm sat}O_j}+\vv{\Omega}\times \left(\vv{\Omega} \times \vv{O_{\rm sat}O_j} \right)$ is the kinematic acceleration. 

Tests of the WEP usually present their result in terms of the E\"otv\"os ratio \cite{eotvos22, will14} : 
\begin{equation} \label{eq_eotvos}
\eta(2,1) =  2 \frac{m_{\rm{G_2}}/m_{\rm{I_2}} - m_{\rm{G_1}}/m_{\rm{I_1}}}{m_{\rm{G_2}}/m_{\rm{I_2}} + m_{\rm{G_1}}/m_{\rm{I_1}}}.
\end{equation}
Note that the  E\"otv\"os ratio depends on the pair of materials used for the test. In this paper, we use a good first order approximation of the E\"otv\"os parameter,
\begin{equation}
\delta(2,1) \equiv \frac{m_{\rm{G_2}}}{m_{\rm{I_2}}} - \frac{m_{\rm{G_1}}}{m_{\rm{I_1}}}.
\end{equation}

We define the common-mode (resp. differential-mode) of a given instrumental parameter as the half-sum (resp. half-difference) of this parameter for both test-masses $o^{(1)}$ and $o^{(2)}$, the inner test-mass being $(1)$ and the outer being $(2)$:
\begin{align}
o^{(c)} = \frac {1} {2} (o^{(1)}+o^{(2)}) \\
o^{(d)} =\frac {1} {2} (o^{(1)} - o^{(2)}).
\end{align}
In what follows, an instrumental parameter can be also defined as a matrix of sensitivity or alignment and noted $[o]$. However, for the measured acceleration $\vv{\Gamma}$, applied acceleration $\vv{\gamma}$ or disturbing accelerations $\vv{b}$ vectors, we use the simple difference instead of the half-difference when defining their differential-mode:
\begin{equation} \label{note_eq1}
\vv{a}^{(c)} = \frac {1} {2} (\vv{a}^{(1)}+\vv{a}^{(2)})
\end{equation}
\begin{equation} \label{note_eq2}
\vv{a}^{(d)} = (\vv{a}^{(1)} - \vv{a}^{(2)}).
\end{equation}
Finally, the distance between the test-mass ${(1)}$ and ${(2)}$, $\vv{O_1O_2}$ (and not $\vv{O_2O_1}$), is noted $\vv{\Delta}$ and called ``offcentering''.

\begin{figure*}
\includegraphics[width=0.45 \textwidth]{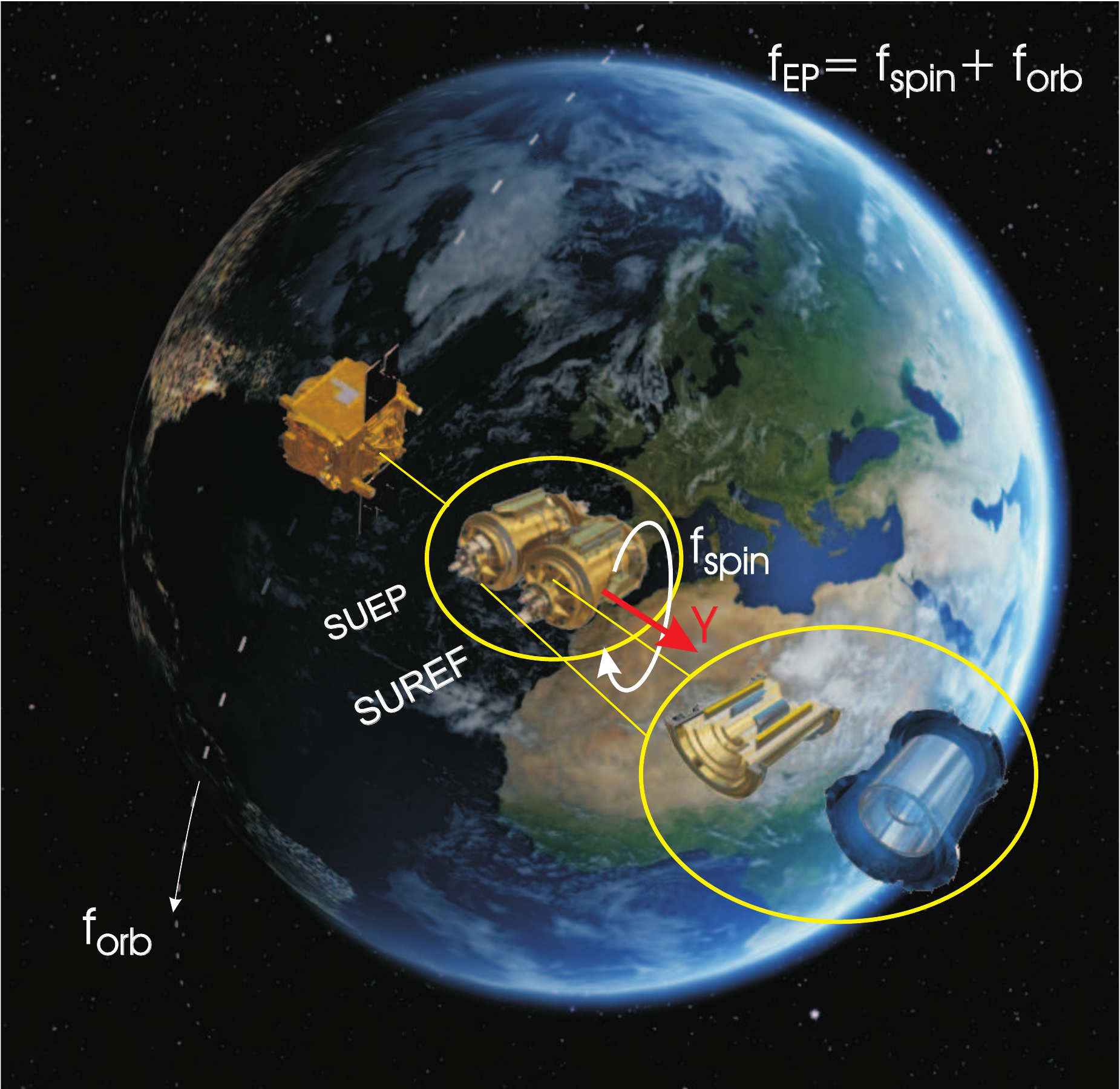}
\includegraphics[width=0.453 \textwidth]{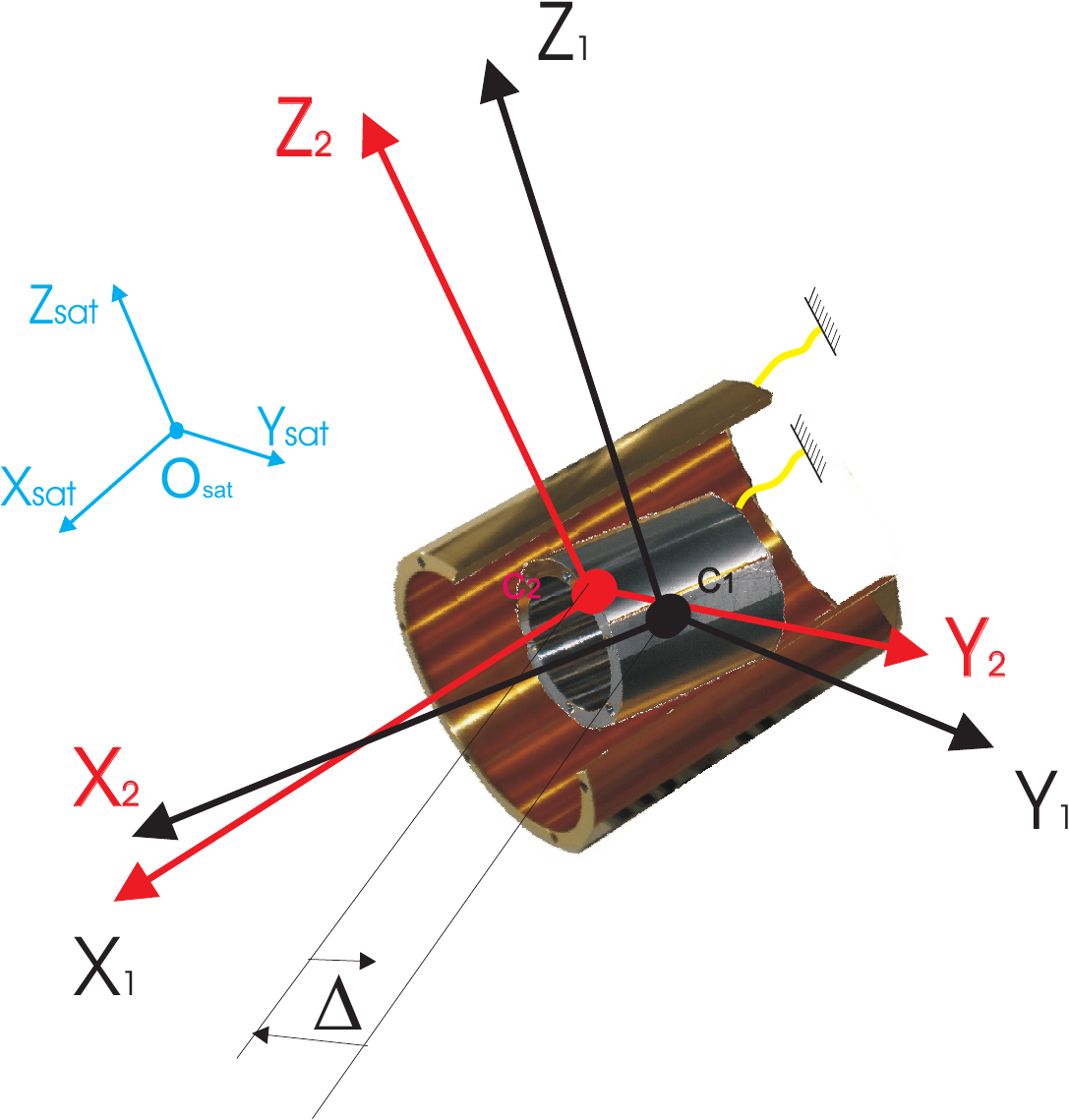}
\caption{Left: the  4 test-masses orbiting around the Earth (credits CNES / Virtual-IT 2017) . Right: test-masses and satellite frames; the ($X_{\rm sat}$, $Y_{\rm sat}$, $Z_{\rm sat}$) triad defines the satellite frame; the reference frames ($X_k$, $Y_k$, $Z_k$, $k=1,2$) are attached to the test-masses (black for the inner mass $k=1$, red for the outer mass $k=2$); the $X_k$ axes are the test-mass cylinders' longitudinal axis and define the direction of Equivalence Principle test measurement; the $Y_k$ axes are normal to the orbital plane, and define the rotation axis when the satellite spins; the $Z_k$ axes complete the triads.  The 7 $\mu$m gold wires connecting the test-masses to the common Invar sole plate are shown as yellow lines. The centres of mass have been approximately identified with the origins of the corresponding sensor-cage-attached reference frames.}
\label{fig_orbit} 
\end{figure*}

The instrument frame is defined in Fig.~\ref{fig_orbit}: the $X$-axis and the $Z$-axis lie in the orbital plane, with the $X$-axis along the main axis of the cylindrical test-masses. The $Y$-axis is normal to the orbital plane, and completes the triad. 

The WEP is measured by monitoring the difference in accelerations undergone by the two test masses along their $X$-axis (the most sensitive). In practice, this is done by computing the difference of measured electrostatic acceleration. In the absence of instrumental defects, the electrostatic acceleration should be in equilibrium with the others actions in Eq. (\ref{eq_meas0}) and lead to a difference of acceleration applied on the test-masses 
\begin{equation} \label{eq_gamma}
\vv\gamma^{(d)} = \delta(2,1) \vv{g}(O_{\rm sat}) + ([{\rm T}] - [{\rm In}]) \vv{\Delta} + \vv{b_1}^{(d)},
\end{equation}
where $[{\rm T}]$ is the Earth gravity gradient tensor projected in the instrument's frame. The kinematic acceleration can be expressed as $- [{\rm In}] \vv{\Delta}$, where $[{\rm In}]$ is the matrix defined by $[{\rm In}] = [\dot \Omega] + [\Omega] [\Omega]$) where $[\Omega]$ is the matrix representation of the angular velocity of the satellite. At last, $\vv{b_1}^{(d)}$ represents the difference of acceleration due to disturbing forces acting on the two test masses in Eq. (\ref{eq_meas0}). 
The electrostatic acceleration applied to each test mass is not perfectly measured, as errors can come from wrong scale of voltage applied to electrodes or from the misalignment of the test-mass measurement frames. Instrumental parameters must therefore be taken into account, such as, for the $j$th test mass:
\begin{itemize}
\item the DC bias (offset) due to the measurement pick-up $\vv{b_0}^{(j)}$;
\item the measurement pick-up noise $\vv{n}^{(j)}$;
\item scale factors (close to 1) $\vv{K_1}^{(j)}=\begin{pmatrix}
K_x^{(j)} & K_y^{(j)} & K_z^{(j)} & K_{\phi}^{(j)}  & K_{\theta}^{(j)} & K_{\psi}^{(j)} 
\end{pmatrix}^t$;
\item couplings $\left[\boldsymbol\eta^{(j)} \right]$ between linear degrees of freedom due to non orthogonality of the measurement frame; this defect is linked to the test-mass geometry and is much smaller than 1; note also that the $\left[\boldsymbol\eta^{(j)} \right]$ matrix is symmetric;
\item couplings between linear degrees of freedom due to electrostatic defects \cite{chhuncqg5}:
\begin{equation}
\left[\boldsymbol\beta^{(j)} \right]=\begin{bmatrix} 
0 & \beta_{xy}^{(j)} &\beta_{xz}^{(j)} \\ 
\beta_{yx}^{(j)} & 0 & \beta_{yz}^{(j)} \\
\beta_{zx}^{(j)} & \beta_{zy}^{(j)} & 0 
\end{bmatrix} \approx \begin{bmatrix} 
0 & 0 & 0 \\ 
\beta_{yx}^{(j)} & 0 & \beta_{yz}^{(j)} \\
\beta_{zx}^{(j)} & \beta_{zy}^{(j)} & 0 
\end{bmatrix}+O(10^{-3});
\end{equation}
\item couplings of linear acceleration with angular acceleration $\left[{\mathbf C}^{(j)} \right]$;
\item rotation from the common instrument frame $\mathcal R_{\rm sat}$ linked to the satellite in which the acceleration $\vv{\gamma}^{(j)}$ is modelled to each test mass frame $\mathcal R_j$ (linked to the test mass geometry) where the acceleration is measured. The alignments of the different frames have been measured or evaluated on the ground prior to the launch and can be approximated in the small angle limit ($\rm{<10^{-2}\,rad}$) by 
\begin{equation}
\left[\boldsymbol\Theta^{(j)} \right]=\begin{bmatrix} 
1 & \theta_z^{(j)} & -\theta_y^{(j)} \\ 
-\theta_z^{(j)} & 1 & \theta_x^{(j)} \\
\theta_y^{(j)} & -\theta_x^{(j)} & 1 
\end{bmatrix}.
\end{equation}
\end{itemize}

These items are combined to construct the linear acceleration measurement parameters
\begin{equation}
\left[ {\mathbf M^{(j)}}\right]=\begin{bmatrix} 
K_x^{(j)} & \eta_z^{(j)} +\beta_{xy}^{(j)} & \eta_y^{(j)}+\beta_{xz}^{(j)} \\ 
\eta_z^{(j)}+\beta_{yx}^{(j)} & K_y^{(j)} & \eta_x^{(j)}+\beta_{yz}^{(j)} \\
\eta_y^{(j)}+\beta_{zx}^{(j)} & \eta_x^{(j)}+\beta_{zy}^{(j)} & K_z^{(j)} 
\end{bmatrix}
\end{equation}
and
\begin{equation}
\left[ {\mathbf A^{(j)}}\right]=\left[ {\boldsymbol\Theta^{(j)}}\right] \left[ {\mathbf M^{(j)}}\right].
\end{equation}

The models of the actually measured differential acceleration and the common mode acceleration (up to leading terms) are then
\begin{equation} 
\vv\Gamma^{(d)} = \vv{b_0}^{(d)} + \left[ {\mathbf A^{(c)}}\right] \vv {\gamma}^{(d)} + 2 \left[{\mathbf A^{(d)}} \right] \vv {\gamma}^{(c)} +2 \left[{\mathbf C^{(d)}} \right] \dt{\vv \Omega} + \vv n^{(d)}
\label{eq_diffmode}
\end{equation} 
and
\begin{equation} \label{eq_commode}
\vv\Gamma^{(c)} = \vv{b_0}^{(c)} + \frac{1}{2} \left[ {\mathbf A^{(d)}}\right] \vv {\gamma}^{(d)} + \left[{\mathbf A^{(c)}} \right] \vv {\gamma}^{(c)} + \left[{\mathbf C^{(c)}} \right] \dt{\vv \Omega} + \vv n^{(c)}.
 \end{equation} 
 
The differential acceleration measurement is used to estimate the E\"otv\"os parameter $\delta (2,1)$ \cite{hardycqg6, bergecqg7}. Coupling effects are negligible on the $X$-axis, in particular the two terms $\eta_z^{(j)} +\beta_{xy}^{(j)}$ and $\eta_y^{(j)}+\beta_{xz}^{(j)}$ that have been estimated in flight \cite{chhuncqg5}. It remains that the measurement can be disturbed by the mismatching of scale factors and the misalignment of the test-masses leading to the projection of the common mode acceleration on $\vv\Gamma^{(d)}$. In order to correct for this defect in (\ref{eq_diffmode}), the applied acceleration $\vv{\gamma}^{(c)}$ is estimated by using the measurement equation (\ref{eq_commode}). This leads to:
\begin{multline} 
\vv\Gamma^{(d)} = \vv{B_0}^{(d)}+ \left[ {\mathbf {\tilde{A}}^{(c)}}\right] \vv {\gamma}^{(d)} +2 \left[{\mathbf A^{(d)}} \right] \left[{\mathbf A^{(c)}} \right]^{-1} \vv {\tilde{\Gamma}}^{(c)}+2\left[{\mathbf C'^{(d)}} \right]  \dt{\vv \Omega} + \vv n^{(d)}
\label{eq_diffT}
 \end{multline} 
where 
\begin{align}
\vv {\tilde{\Gamma}}^{(c)}&=\vv {\Gamma}^{(c)}-\vv n^{(c)}, \\
\vv{B_0}^{(d)}&= \vv{b_0}^{(d)} -2 \left[{\mathbf A^{(d)}} \right] \left[{\mathbf A^{(c)}} \right]^{-1} \vv{b_0}^{(c)}, \\
\left[{\mathbf C'^{(d)}} \right]&=(\left[{\mathbf C^{(d)}} \right]- \left[{\mathbf A^{(d)}} \right] \left[{\mathbf A^{(c)}} \right]^{-1} \left[{\mathbf C^{(c)}} \right] ) \\
\left[ {\mathbf {\tilde A}^{(c)}}\right] &= (\left[ {\mathbf A^{(c)}}\right] - \left[{\mathbf A^{(d)}} \right] \left[{\mathbf A^{(c)}} \right]^{-1}\left[{\mathbf A^{(d)}} \right] ) \approx \left[ {\mathbf {A}^{(c)}}\right] ~ {\rm at~first ~order}.
\end{align}

%
%
%

The DFACS applies a thrust which reduces significantly the mean output $\vv {\Gamma}^{(c)}$. More precisely, most of the sessions were performed with the DFACS controlled on the outer test-mass, some on the inner test-mass and very few on the mean of the two test-masses. In addition, in order to minimise the gas consumption, the DFACS does not compensate the DC bias of the accelerometer. The accelerometer bias is regularly estimated as $\vv{\hat{b}_0}^{(c)}$ and injected in the DFACS loop to subtract the DC bias in the thrust command \cite{robertcqg3, rodriguescqg4}. As a result, the common mode measurement (or the test-mass reference for the DFACS loop acceleration measurement) is not totally nullified and becomes $\vv{\Gamma}^{(c)}=\vv{\hat{b}_0}^{(c)}+\vv{R}_{\rm df}$, where $\vv{\hat{b}_0}^{(c)}$ is the estimation of $\vv{{b}_0}^{(c)}$ and $\vv{R}_{\rm df}$ the residual error control of the DFACS seen by the accelerometer (i.e. it includes all alignment, coupling and scale factor errors as they are all compensated by the closed loop).

The WEP is tested along the $X$-axis and the equations are expressed in the ${\mathcal R}_{SU}$ frame defined by the test-mass used for the DFACS (or the mean frame of the two test-masses when using the mean of both). The equations in this paper are expressed in the mean frame  $\frac{1}{2}({\mathcal R}_{(1)}+{\mathcal R}_{(2)})$. So the measurement is the projection of Eq. (\ref{eq_diffT}) on the $X$-axis, after substituting Eq. (\ref{eq_gamma}) in Eq. (\ref{eq_diffT}):

\begin{equation}  \label{eq_xacc}
\begin{split}
\Gamma_x^{(d)} &\approx  B_{0x}^{(d)} \\
    &+  \tilde{a}_{c11} b_{1x}^{(d)} + \tilde{a}_{c12} b_{1y}^{(d)}+ \tilde{a}_{c13} b_{1z}^{(d)} \\
    &+ \tilde{a}_{c11} \delta g_x + \tilde{a}_{c12} \delta g_y + \tilde{a}_{c13} \delta g_z \\
    & + \left(T_{xx} - In_{xx} \right) \tilde{a}_{c11} \Delta_x + \left(T_{xy} - In_{xy} \right) \tilde{a}_{c11} \Delta_y + \left(T_{xz} - In_{xz} \right) \tilde{a}_{c11} \Delta_z \\
    & +  \left(T_{yx} - In_{yx} \right) \tilde{a}_{c12} \Delta_x + \left(T_{yy} - In_{yy} \right) \tilde{a}_{c12} \Delta_y + \left(T_{yz} - In_{yz} \right) \tilde{a}_{c12} \Delta_z \\
    & + \left(T_{zx} - In_{zx} \right) \tilde{a}_{c13} \Delta_x + \left(T_{zy} - In_{zy} \right) \tilde{a}_{c13} \Delta_y  + \left(T_{zz} - In_{zz} \right) \tilde{a}_{c13} \Delta_z \\
    &+ 2\left( \frac{a_{d11}}{a_{c11}} {\Gamma}_x^{(c)} + \frac{a_{d12}}{a_{c22}}{\Gamma}_y^{(c)} + \frac{a_{d13}}{a_{c33}} {\Gamma}_z^{(c)}  \right)\\ 
    &+ 2 \left(c'_{d11} \dt{\Omega}_x + c'_{d12} \dt{\Omega}_y + c'_{d13} \dt{\Omega}_z  \right)\\
    &+ n_x^{(d)}-2\left( \frac{a_{d11}}{a_{c11}} n_x^{(c)} + \frac{a_{d12}}{a_{c22}} n_y^{(c)} + \frac{a_{d13}}{a_{c33}} n_z^{(c)}  \right) ,\\
\end{split}
\end{equation}
where
\begin{itemize}
\item $\left[a_{c11}, a_{c12}, a_{c13} \right]$ is the first line of the common-mode sensitivity matrix,
\item $\left[a_{d11}, a_{d12}, a_{d13} \right]$ is the first line of the differential-mode sensitivity matrix $\left[ {\mathbf A^{(d)}}\right] $,
\item $\left[c'_{d11}, c'_{d12}, c'_{d13} \right]$  is the first line of the differential-mode angular to linear coupling matrix $\left[ {\mathbf C'^{(d)}}\right] $,
\end{itemize}

The first line of Eq. (\ref{eq_xacc}) is the effect of the bias in differential and common mode mostly at DC, which can drift with time. The second line is the projection onto the sensor frame of the effect of the parasitic differential accelerations acting on the test-masses with contributions at DC and $f_{\rm{EP}}$ (systematic errors are detailed in Ref. \cite{hardycqg6}). The third line is the projection of the $\delta \vv g$ term on the mean $X$-axis of the two test-masses. As $a_{c12}$ and $a_{c13}$ are close to $10^{-3}~{\rm rad}$, we consider only  $a_{c11} \delta g_x$. The 4th, 5th and 6th lines describe the effect of the test-mass offcentering coupled to the Earth gravity gradient and the inertia motion projected on the mean $X$-axis of the two test-masses. The 7th line describes the effect of the common mode acceleration due to the mismatching of scale factors or alignments. The 8th line represents the coupling of angular motion on the linear acceleration measurement along the $X$-axis. The last line is the measurement noise contribution.


The calibration process allows us to determine the $a_{c11} \Delta_x$, $a_{c11} \Delta_y$, $a_{c11} \Delta_z$, $\frac{a_{d11}}{a_{c11}}$, $\frac{a_{d12}}{a_{c11}}$ and $\frac{a_{d13}}{a_{c11}}$ terms, such that the calibrated acceleration is given by :
\begin{equation}  \label{eq_xacc_cor}
\begin{split}
\Gamma_x^{(d,{\rm cal})} =& \left(T_{xx} - In_{xx} \right)^{\rm mod}[a_{c11} \Delta_x]^{\rm cal}+ \left(T_{xy} - In_{xy} \right)^{\rm mod}[a_{c11} \Delta_y]^{\rm cal}\\
        &+ \left(T_{xz} - In_{xz} \right)^{\rm mod}[a_{c11} \Delta_z]^{\rm cal} \\
        &+ 2\left( \left[ \frac{a_{d11}}{a_{c11}}\right]^{\rm cal} {\Gamma}_x^{(c)} + \left[\frac{a_{d12}}{a_{c22}} \right]^{cal}{\Gamma}_y^{(c)} + \left[\frac{a_{d13}}{a_{c33}} \right]^{cal} {\Gamma}_z^{(c)}  \right) \\
        &-2\left( \left[\frac{a_{d11}}{a_{c11}}\right]^{\rm cal} n_x^{(c)} +\left[\frac{a_{d12}}{a_{c22}}\right]^{\rm cal} n_y^{(c)} + \left [\frac{a_{d13}}{a_{c33}}\right]^{\rm cal} n_z^{(c)}  \right), \\
\end{split}
\end{equation}
which leads to the corrected measurement equation
\begin{equation}  \label{eq_xacc_cr}
\Gamma_x^{(d, {\rm cor})} = \Gamma_x^{(d)} - \Gamma_x^{(d, {\rm cal})}.
\end{equation}


\section{Requirement tree and expected budget performance} \label{sect_req}

Requirements are established by considering all errors that could bias the measurements (Eqs. \ref{eq_xacc} and \ref{eq_xacc_cr}). As the looked-for signal is at well known frequency, aliasing phenomena are also taken into account in the requirement tree.

We present in this paper two budgets of performance: 
\begin{itemize}
\item one prior to the launch, with inputs coming from the requirements on all satellite subsystems for a test in V2 mode;
\item an update of this budget established after the flight commissioning phase, that takes into account new configurations of the satellite (V3 mode) and of the instrument's servo-control.
\end{itemize}
The following subsections focus on the instrument's $X$-axis. A similar analysis was performed on the other five degrees of freedom. The derived requirements are less stringent than on $X$-axis.
Ref. \cite {hardycqg6} gives the performance obtained with actually measured inputs and should be compared to the error budget presented here.

\subsection{Frequencies of interest} \label{sect:freq}

Depending on the satellite mode, several frequencies of interest have been used to check the error budget. 
Independently of the mode, requirements are established at the frequency $f_{\rm{EP}}$ of a potential WEP violation signal and at twice this frequency, $2f_{\rm{EP}}$ corresponding to the main signal due to the Earth gravity gradient that allows for the estimation of $a_{c11} \Delta_x$ and  $a_{c11} \Delta_z$. Additionally, $3f_{\rm{EP}}$ has also been considered to establish the requirements on the system to limit their projection effect at $f_{\rm{EP}}$ through non linearities. Finally, a reference signal at $f_{\rm{cal}}$ is used during calibration sessions in inertial pointing, so that signals at $f_{\rm{cal}}$ and $2f_{\rm{cal}}$ have also been taken into consideration.

Ref. \cite{hardy13b} shows how a signal at any frequency may be projected at $f_{\rm{EP}}$ because of the finite duration of the observation window, hereby perturbing the measurement. This leads to the definition of a pattern of the projection rate at  $f_{\rm{EP}}$ of a disturbing signal at a different frequency. Fig. \ref{fig_pattern} shows the shape of the pattern determined to reject a signal at a level of $2 \times{}10^{-16}$\,m\,s$^{-2}$ which is the error allocation of one error source at $f_{\rm{EP}}$ (see below for error allocation).

\begin{minipage}{0.5\textwidth} 
\includegraphics[scale=0.3]{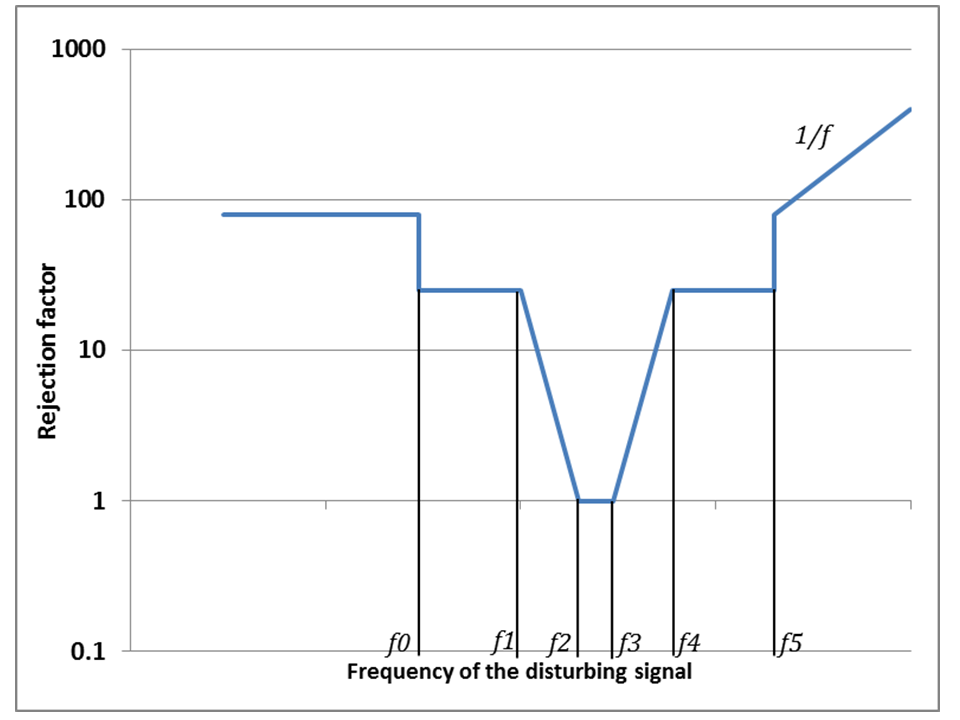}
\captionof {figure} {Required rejection rate of disturbing signal versus frequency}
\label{fig_pattern} 
\end{minipage}
\hspace{2ex}
\begin{minipage} {0.5\textwidth}
\begin{tabular}{ll}
     \br
     Frequency & Value \\
     \mr
     $f_0$ & 0 in inertial pointing \\
      & or ($f_{\rm{EP}}-2f_{\rm{orb}}$) otherwise\\
     \mr
     $f_1$ & $f_{\rm{EP}}-f_{\rm{orb}}/2$\\
     \mr
     $f_2$ & $f_{\rm{EP}}-f_{\rm{orb}}/20$\\
     \mr
     $f_3$ & $f_{\rm{EP}}+f_{\rm{orb}}/20$\\
     \mr
     $f_4$ & $f_{\rm{EP}}+f_{\rm{orb}}/2$\\
     \mr
     $f_5$ & $f_{\rm{EP}}+2f_{\rm{orb}}$\\
     \br
     \\
     \\
\end{tabular}
\end{minipage}
The spin frequencies are fixed so that  $f_{\rm{spin}}=\frac{k}{2} f_{\rm{orb}}$, $k$ being an integer ($k=9$ for $f_{\rm{spin}_1}$ and $k=35$ for $f_{\rm{spin}_2})$. Moreover the duration $T_{ss}$ of the  science sessions are fixed to an even number of orbital periods, $T_{ss}=2nT_{\rm{orb}}$. This ensures that the session contains integer numbers of orbital periods, of spin periods, and of EP periods:
\begin{equation}
  \label{eq:1}
  T_{ss}=2nT_{\rm orb}=nkT_{\rm spin}=n(k+2)T_{\rm EP}
\end{equation}
The benefit is a natural decorrelation between periodic signals at frequencies multiples of $f_{\rm{orb}}$, $f_{\rm{spin}}$ and $f_{\rm{EP}}$.
Calibration sessions are less demanding; in order to ensure an integer number of calibration periods, their duration has been set to 5.07 orbits (see for detailed list of sessions in Ref. \cite {rodriguescqg4}).

\subsection{Requirement tree} \label{sect:reqt}

The specification of all error sources have been determined by considering four types of errors:
\begin{itemize}
\item harmonic errors $S_{\rm{EP}}$ at $f_{\rm{EP}}$ in phase with the gravity acceleration due to the Earth, source of a possible signal of violation;
\item errors $S(f')$ at frequencies $f'$ around $f_{\rm{EP}}$ that project to the frequency of interest with a rejection rate $R'(f')$;
\item harmonic errors $S(n f_{\rm{EP}})$, if $n=1$ in quadrature with the signal we seek,  for $n\ne1$ at the $n$ multiple $f_{\rm{EP}}$ frequency, the error is rejected by a factor $R(n f_{\rm{EP}}$);
\item stochastic error $S_{\rm{r}}$ around $f_{\rm{EP}}$.
\end{itemize}

To reach the accuracy of $10^{-15}$, a signal-to-noise ratio of 1 was considered, implying that the total error of the accelerometric measurement must be lower than $7.9\times{}10^{-15}$\,m\,s$^{-2}$ (since the Earth gravity acceleration amplitude at 710\,km is 7.9\,m\,s$^{-2}$).
All errors have been defined in the frequency domain as a discrete Fourier transform for the errors at the interest frequencies and as a power spectral density for random errors. Harmonic errors are added which is a conservative approach since some of the errors are not correlated and could be added quadratically. The case of $S(n f_{\rm{EP}})_{n=1}$ has been considered in the same way as $S_{\rm{EP}}$ to establish the requirements. The budget error considered the total error at $f_{\rm{EP}}$  as $S_{\rm{T}}$ in the following way: 
\begin{equation}  \label{tot_err}
    S_{\rm{T}}^2= \left[\sum{S_{\rm{EP}}}+\sum {\frac{S(f')}{R(f')}} + \sum {\frac{S(n f_{\rm{EP}})}{R(n f_{\rm{EP}})}}\right]^2+\sum {\frac{S_{\rm{r}}^2}{T_{\rm{ss}}}}.
\end{equation}

The total allocation error for $S_{\rm{T}}$ is $7.9\times{}10^{-15}$\,m\,s$^{-2}$ distributed equally between harmonic and random errors to $(7.9\times{}10^{-15})/\sqrt{2}$\,m\,s$^{-2}$. This leads to allocate the following budget over 120 orbits:
\begin{align}
 \sqrt{\sum {S_{\rm{r}}^2} } &< 4.7\times{}10^{-12}~{\rm m\,s^{-2}Hz^{-1/2}}, \\
 \left[\sum{S_{\rm{EP}}}+\sum {\frac{S(f')}{R(f')}} + \sum {\frac{ S(i f_{\rm{EP}})}{R(i f_{\rm{EP}})}}\right] &<5.6\times{}10^{-15}~{\rm m\,s^{-2}}.
\end{align}

From this global allocation, the number of budget allocations for random error sources and of systematic errors has been set as follows: 200 allocations for random sources at a level of $3\times{}10^{-13}$\,m\,s$^{-2}$\,Hz$^{-1/2}$ each and 28 for systematic sources at a level of $2\times{}10^{-16}$\,m\,s$^{-2}$ each. From Eq. (\ref{eq_xacc_cr}), the main error sources have been evaluated in order to distribute allocations of errors to the subsystems. After several iterative analyses between the satellite and the instrument team, a final budget is realised for the main error sources :

\begin{itemize}
\item the instrument measurement noise with a particular allocation of $3.3\times{}10^{-12}$\,m\,s$^{-2}$\,Hz$^{-1/2}$ in differential mode leaving a budget of $3.4\times{}10^{-12}$\,m\,s$^{-2}$\,Hz$^{-1/2}$ for all the other stochastic noises measured in the difference of acceleration;
\item the instrument bias sensitivity to environment (random or systematics). That includes sensitivity to fluctuations of magnetic field, of local gravity and of temperature. As the thermal sensitivity is a major constraint in the design, a larger allocation was fixed to $10^{-15}$\,m\,s$^{-2}$;  
\item the instrument parameters $\left[{\mathbf A^{(j)}} \right]$ and $\left[{\mathbf C^{(j)}} \right]$  variations (noise or systematics) due to temperature fluctuations;
\item the satellite position knowledge accuracy that could lead to systematic errors in the evaluation of the gravity field and its gradients;
\item the satellite orientation  knowledge accuracy that could lead to systematic errors in the evaluation of the gravity field phase that determines the phase of the possible WEP violation signal;
\item the satellite angular velocity and acceleration noise and systematics; 
\item the drag-free control of the linear common mode acceleration (noise and systematics);
\item the dating of the measurement data.
\end{itemize}

\subsection {Required performance budget}

The pre-launch baseline configuration was the spin V2 mode.
Table \ref{tab_budgetV2} shows the corresponding expected performance when considering all the error sources.

The sources of error may vary randomly or at the EP frequency. For example, in science mode, with the DFACS turned on, Eq.(\ref{eq_xacc})'s term $ \frac{a_{d11}}{a_{c11}} {\Gamma}_x^{(c)}=\frac{a_{d11}}{a_{c11}} \left({\hat{b}_0}_x^{(c)}+R_{\rm{df}_x}\right)$ can vary with time because the term $ \frac{a_{d11}}{a_{c11}}$  varies randomly (electronic noise) or  systematically (temperature variations), and because the drag-free performance $R_{\rm{df}_x}$ varies randomly (gas thruster noise) or systematically (star sensor harmonic errors). Depending on the origin of the disturbance, the errors are classified in different topics on the tables: Earth gravity gradients, instrument gravity, angular motions, temperature variations, etc. The drag-free performance expressed by $R_{\rm{df}_x}$ is listed in the Drag-free residuals. The accelerometer noise and parasitic forces comprise the electronic measurement noise, the effect of the temperature gradients (radiometer effect, radiation pressure), the gold wire stiffness and damping and the contact potential difference impact.

\begin{table}
\caption{\label{tab_budgetV2} Expected a priori budget of performance in V2 mode.}
\begin{indented}
\item[]\begin{tabular}{@{}lll}
     \br
     \textbf{Error source} & \textbf{Contribution in} & \textbf{Contribution in}  \\
      SUREF in V2     & \textbf{Random noise} & \textbf{Harmonic error}\\  
                        & m\,s$^{-2}$\,Hz$^{-1/2}$ & m\,s$^{-2}$\\
     \mr
   Earth gravity gradients & $0.0\times{}10^{-13}$ & $0.6\times{}10^{-16}$ \\
     \mr
    Instrument gravity & $0.6\times{}10^{-13}$ & $2.0\times{}10^{-16}$ \\
     \mr
    Satellite gravity gradients & $1.2\times{}10^{-13}$ & $3.1\times{}10^{-16}$\\
     \mr
    Angular motions &$3.8\times{}10^{-13}$ & $11.0\times{}10^{-16}$\\
     \mr
    Instrument parameter variations &$6.0\times{}10^{-13}$ & $8.5\times{}10^{-16}$\\
     \mr
    Accelerometer measurement noise & $14.6\times{}10^{-13}$ & \\
    and parasitic forces                      &&\\
     \mr
    Temperature variations & $2.1\times{}10^{-13}$ & $8.6\times{}10^{-16}$\\
     \mr
    Drag-Free Residuals & $1.0\times{}10^{-13}$ & $5.0\times{}10^{-16}$\\
     \mr
    Magnetic sensitivity & $0.8\times{}10^{-13}$ & $4.0\times{}10^{-16}$\\
     \mr
    Non linearity & $3.7\times{}10^{-13}$ & $8.1\times{}10^{-16}$\\
     \br
   Total & $16.9\times{}10^{-13}$ & $55.9\times{}10^{-16}$\\
     \br
   \textbf{EP test budgeted for $\delta : 0.7\times{}10^{-15}$}& with $g=7.9$\,m/s$^2$& \\
    &over 120 orbits&\\
     \br
\end{tabular}
\end{indented}
\end{table}

The driving rules to establish each term of the table are detailed below.

\subsection{Gravity field signal} \label{sect:grav}

As they determine of the phase of $\vv{g}$ in Eq. (\ref{eq_xacc}), the date, the position and the orientation of the satellite with respect to the Earth must be measured. We use the method detailed in Ref. \cite{metris98} to compute the Earth gravity acceleration and its gradient tensor projected onto the instrument frame using the ITSG-Grace2014s gravity potential model \cite{mayer06} expanded up to spherical harmonic degree and order 50.  
 
The position of the satellite is obtained from the Doppler telemetry measurements associated to the on-board GPS data \cite{robertcqg3}. Table \ref{tab_satpos} gives the requirements on the satellite's position to limit the error of estimation on $T_{xx}$ and $T_{xz}$ at $f_{\rm{EP}}$.
The offcentering is specified to 20$\mu$m on $\Delta x$ and $\Delta z$ with a knowledge accuracy of 0.1$\mu$m and to 20$\mu$m on $\Delta y$ with a knowledge accuracy of 2$\mu$m. The pointing stability and knowledge has to be also accurate to limit the effect of the gravity gradients. 

The altitude of the satellite determines the intensity of the gravity field and has been specified to 710\,km as a compromise between the maximization of the signal magnitude and the minimization of the external forces on the satellite: the Solar radiation pressure and drag forces are of the same order of magnitude at about 700\,km. The final figure was fixed in agreement with the main passenger of the Soyutz launch. The satellite has to follow a (near polar) sun-synchronous orbit to maximise the thermal stability. The orbit's eccentricity must be lower than $5\times{}10^{-3}$ to limit the amplitude of the gravity gradient components at $f_{\rm{EP}}$.
  
\begin{table}
\caption{\label{tab_satpos} Satellite position knowledge and pointing accuracy requirements}
\begin{indented}
\item[]\begin{tabular}{@{}lrrrl}
     \br
     \textbf{Frequency} & \textbf{Radial} & \textbf{Tangent}  & \textbf{Cross track} & \textbf{Orientation}  \\
     \mr
   DC & 100\,m & 100\,m & 2\,m & 2.5$\times{}10^{-3}$\,rad \cr
        &             &             &        & known at $10^{-3}$\,rad \cr
     \mr
    $f_{\rm{EP}}$ & 7\,m & 14\,m& 100\,m& 10\,$\mu$rad \cr
        &             &             &        &known at 1\,$\mu$rad \cr
     \mr
    2$f_{\rm{EP}}$ & 100\,m & 100\,m& 2\,m& not stringent\cr
     \mr
    3$f_{\rm{EP}}$ & 2\,m & 2\,m& 100\,m&10\,$\mu$rad \cr
        &             &             &        &known at 1\,$\mu$rad\cr
     \br
\end{tabular}
\end{indented}
\end{table} 

\subsection {Instrument and satellite self gravity} \label{sect:selfgrav}

The gravity field generated by local moving test-masses could also mimic a gravity signal at $f_{\rm{EP}}$. As the test-masses cannot be considered as point masses for local gravity, the difference of gravity exerted on the test-masses depends on their shape. Refs \cite{Lockerbie93, willemenot97} show that for a cylindrical test mass, the gravitational potential produced by a moving mass $M_s$ at the distance $R_s$ making an angle $\theta$ with the cylinder axis can be expressed as a sum of Legendre  polynomials $P_{2p}$ with form factors $k_{2p}$ depending only on geometry:
\begin{equation}
V(R_s)=-\frac{G M_s}{R_s} \sum_{p=0}^{\infty}k_{2p} \frac{P_{2p}(cos\theta)}{R_s^{2p}},
\end{equation}

Order 0 is the point source case. When looking for the difference of gravity field $\Delta g$ exerted on two concentric test-masses by the moving mass, order 0 disappears and it remains only order 1 depending on the test-mass's moments of inertia:
\begin{equation}
\Delta g \approx 12 \frac {G M_s}{R_s^4} \left(\frac{J}{m}\right)_{\rm max}\frac{\Delta J}{J},
\end{equation}
where $\Delta J/J$ represents the mean difference $(J_i-J_j)_{i=x,y,z; j \neq i}$ of the test-mass' moments of inertia about $X$, $Y$ and $Z$. If all moments of inertia are identical about the three degrees of freedom, then the test-mass can be considered point-like and the sensitivity to local moving mass is reduced. For a perfect hollow cylinder of inner radius $R_1$ and outer radius $R_2$, if the length of the test-mass is defined as $L=\sqrt{\frac{3}{4}(R_2^2+R_1^2)}$, then the moments of inertia are equal along the three axes at second order in $R_i^2$. As test-masses have some flat parts and holes, this relation must be adjusted by computing the moment of inertia of the real shape. The ratio $\Delta J/J$ can also vary with the geometry accuracy or with the density inhomogeneity. A specification on each error contributor has been set to $2\times{}10^{-4}$.

Thermal dilation of the satellite and of the instrument changes their mass distribution and thus the local gravity field, impacting the dynamics of test-masses.
Specifications on the self-gravity at 2$f_{\rm{EP}}$ are driven by the required precision of the offcentering estimation (performed with the component of the Earth's gravity gradient at 2$f_{\rm{EP}}$). The acceleration obtained with the self-gravity coupled to the offcentering of 20$\mu$m must be lower than the acceleration residual of the Earth's gravity gradient effect once the offcentering is estimated at a precision of 0.1$\mu$m. The main specifications are presented in Table~\ref{tab_selfgrav}.

\begin{table}
\caption{\label{tab_selfgrav} Main requirements on self gravity induced by the satellite and the instrument.}
\begin{indented}
\item[]\begin{tabular}{@{}lrl}
     \br
     \textbf{Component} & \textbf{Specification} & \textbf{Constraining source}  \cr
     \mr
   DC gravity & $5\times{}10^{-9}$\,m\,s$^{-2}$ & Mass distribution\cr
     \mr
   Differential acceleration at $f_{\rm{EP}}$ & 10$^{-16}$\,m\,s$^{-2}$ & Test-mass shape \cr
     \mr
   Common mode acceleration at $f_{\rm{EP}}$ & 10$^{-12}$\,m\,s$^{-2}$ & Thermal expansion \cr
     \mr
        Common mode acceleration at 2$f_{\rm{EP}}$ & $2\times{}10^{-11}$\,m\,s$^{-2}$ & Thermal expansion \cr
     \mr
    Gravity gradients at $f_{\rm{EP}}$ & $5\times{}10^{-12}$\,s$^{-2}$  & Mass distribution \cr
     &   & with thermal expansion\cr
     \mr
    Gravity gradients at 2$f_{\rm{EP}}$ & $5\times{}10^{-10}$\,s$^{-2}$  & Mass distribution \cr
     &   & with thermal expansion\cr
     \br
\end{tabular}
\end{indented}
\end{table}

\subsection{Angular motion} \label{sect:ang}

Angular motion is dominated by centrifugal and Coriolis effects defined by the matrix $[{\rm In}]$. The 20$\mu$m offcentering specification is considered as baseline.
At first order the first line of $\left[ {\mathbf A^{(c)}}\right] $  can be approximated by $\left[1+O(10^{-2}), O(10^{-3}), O(10^{-3}) \right]$. The differential linear acceleration induced by the angular velocity and the angular acceleration thus reads:  
\begin{equation}  \label{ang_err}
   (-\Omega_{y}^2-\Omega_{z}^2)\Delta_{x}+(\Omega_x \Omega_y-\dot{\Omega}_z)\Delta_y+(\Omega_x \Omega_z+\dot{\Omega}_y)\Delta_z.
\end{equation}

A precise estimation of the angular motion is estimated a posteriori \cite{robertcqg3} which helps to estimate this error contribution. It can be also corrected in the data science process \cite{bergecqg8}.

To derive the requirements from Eq. (\ref{ang_err}), we note that $[\Omega]$ and $[\dot\Omega]$ are distributed over several frequency components: $[\Omega]=[\Omega]^{0}+[\Omega]^{f_{\rm{EP}}}+[\Omega]^{2f_{\rm{EP}}}+[\Omega]^{3f_{\rm{EP}}}$, since $[\Omega]$ is squared in the equation. When spinning the satellite, $\Omega_y=2\pi f_{\rm{spin}}$ is much larger than the angular velocity residual in inertial pointing. This last term determines the specification at $f_{\rm{EP}}$, then the specification at $2f_{\rm{EP}}$ and $3f_{\rm{EP}}$ are derived.

Requirements have been established for all satellite modes. Table \ref{tab_angreq} summarises the main specifications in V2 rotating mode. During calibration sessions, the specifications at $f_{\rm{cal}}$, 2$f_{\rm{cal}}$ and 3$f_{\rm{cal}}$ are relaxed by two to three orders of magnitude compared to the ones at $f_{\rm{EP}}$, 2$f_{\rm{EP}}$ and 3$f_{\rm{EP}}$ as the sensitivity needed in the differential acceleration measurement is about a few 10$^{-12}$\,m\,s$^{-2}$, three orders of magnitudes higher than the WEP test sensitivity target.
Pointing is also specified to prevent a projection of transverse component of the gravity gradients at $f_{\rm{EP}}$ and 2$f_{\rm{EP}}$.

\begin{table}
\caption{\label{tab_angreq} Main requirements on satellite angular motion in V2 mode.}
\begin{indented}
\item[]\begin{tabular}{@{}lrl}
     \br
     \textbf{Component} & \textbf{Specification} & \textbf{In instrument frame}  \cr
     \mr
   Sat. pointing &  &  \cr
   at $f_{\rm{EP}}$ and $3f_{\rm{EP}}$ & $10\mu$rad & all axes \cr
 A posteriori knowledge  & $1\mu$rad & all axes \cr
     \mr
   Angular velocity &  &  \cr
   at $f_{\rm{EP}}$ & $5\times{}10^{-9}$\,rad\,s$^{-1}$ & $\Omega_x$ and $\Omega_y$ \cr
    at $f_{\rm{EP}}$  & $3.5\times{}10^{-7}$\,rad\,s$^{-1}$ & $\Omega_z$ \cr
    at 2$f_{\rm{EP}}$ and 3$f_{\rm{EP}}$ & $3.5\times{}10^{-7}$\,rad\,s$^{-1}$ & all axes \cr
     \mr
   Angular acceleration &  &  \cr
    at $f_{\rm{EP}}$   & $5\times{}10^{-12}$\,rad\,s$^{-2}$ & all axes \cr
    at 2$f_{\rm{EP}}$ and 3$f_{\rm{EP}}$ & $10^{-9}$\,rad\,s$^{-2}$ & all axes \cr
     \br
\end{tabular}
\end{indented}
\end{table}

\subsection{Instrumental parameters} \label{sect:scale}

This subsection applies to the effect of temporal variations of instrumental parameters $[A^{(c)}]$ and $[A^{(d)}]$. The temperature stability of the instrument parameters is discussed in section \ref{sect:therm}. The expression of the disturbance can be summarised as
$[nA^{(c)}]\gamma^{(d)}+[nA^{(d)}]\gamma^{(c)}$, where $[nA^{(j)}]$ is the stochastic noise of the matrix $[A^{(j)}]$.
Table \ref{tab_nAc} lists the main requirements on the instrument's parameters stability under some hypotheses on $\gamma^{(d)}$ and $\gamma^{(c)}$. For $\gamma^{(c)}$, a mean acceleration of $3\times{}10^{-8}$\,m\,s$^{-2}$ has been specified. This acceleration is the residual acceleration applied to the satellite,  the DC bias of the accelerometer being subtracted from the thruster command. For the mean differential acceleration $\gamma^{(d)}$, the DC gravity gradient or angular motion effects have a negligible effect coupled to the stability of alignments. For the scale factor stability, $\gamma^{(d)}$ is dominated by the part of the differential bias sensitive to the scale factor variation on $X$: $2.5\times{}10^{-8}$\,m\,s$^{-2}$.

\begin{table}
\caption{\label{tab_nAc} Main requirements on instrument's parameters.}
\begin{indented}
\item[]\begin{tabular}{@{}lllcl}
     \br
     \textbf{Component} & \textbf{Common} & \textbf{Differential} & SU & FEEU \cr
       & \textbf{noise} & \textbf{noise} & \multicolumn{2}{c}{Temp. sensitivity} \cr
     \mr
   Scale factor & $10^{-5}$\,Hz$^{-1/2}$ & $3\times{}10^{-6}$\,Hz$^{-1/2}$ &$8\times{}10^{-6}$\,K$^{-1}$& $2\times{}10^{-6}$\,K$^{-1}$\cr
     \mr
   Alignment & $10^{-5}$\,rad\,Hz$^{-1/2}$ & $10^{-7}$\,rad\,Hz$^{-1/2}$ &$10^{-7}$\,K$^{-1}$& $2\times{}10^{-8}$\,K$^{-1}$\cr
     \mr
 Lin. Coupling & \multicolumn{2}{c}{$10^{-7}$\,rad\,Hz$^{-1/2}$}&$5\times{}10^{-8}$\,K$^{-1}$& $6\times{}10^{-9}$\,K$^{-1}$\cr
     \mr
 Ang. Coupling & \multicolumn{2}{c}{$10^{-8}$\,(m\,s$^{-2}$)/(\,rad\,s$^{-2}$)\,Hz$^{-1/2}$ }&$5\times{}10^{-9}$\,K$^{-1}$& $6\times{}10^{-10}$\,K$^{-1}$\cr
     \br
\end{tabular}
\end{indented}
\end{table}

\subsection{Accelerometer noise} \label{sect:noise}
This subsection is devoted to the stochastic measurement noise of the accelerometer. Systematic errors are discussed in subsections \ref{sect:selfgrav}, \ref{sect:therm} and \ref{sect:mag}.
The measurement noise of the accelerometer is driven by several sources \cite{liorzoucqg2, chhuncqg5}:
\begin {itemize}
\item Electrostatic stiffness coupled to the servo loop position noise;
\item Actuation noise of the electronics that applies the voltage to the electrodes in order to create electrostatic forces; this noise includes the digital conversion noise;
\item Contact Potential Difference (CPD) at the surface of electrodes and of test-masses that disturbs the applied voltages, evaluated to less than $10\mu$V\,Hz$^{-1/2}$;
\item The stiffness and mechanical damping of the 7\,$\mu$m gold wire that applies the reference voltage to the test-mass.
\end {itemize}

Random or systematic temperature variations that have an effect on scale factors, alignments, electronics biases and parasitic forces (radiation pressure, radiometer effect, outgassing) are discussed in subsection \ref{sect:therm}.

The dominant term is the gold wire damping on the $X$-axis and the electrostatic stiffness effect and actuation noise for the $Y$ and $Z$ axes.
Prior to the launch, two measurement channels were foreseen for $X$: one science channel with a specification of $2\times{}10^{-12}$\,m\,s$^{-2}$\,Hz$^{-1/2}$ and one DFACS channel with a specification of $10^{-11}$\,m\,s$^{-2}$\,Hz$^{-1/2}$. On the less demanding axes, $Y$ and $Z$, the specification was set to  $10^{-10}$\,m\,s$^{-2}$\,Hz$^{-1/2}$.

After the launch, the science channel on one SUEP test-mass was saturated, which led to a new measurement strategy: the DFACS channels are used for the science, the three remaining science channels being used only for instrument characterisation. The performance presented here is realised on the DFACS utilisation baseline.

\subsection{Thermal sensitivity} \label{sect:therm}
The acceleration measurement is sensitive to the temperature of the SU and of the FEEU.
The sensitivities of $[A^{(c)}]$, $[A^{(d)}]$ and $[C^{(d)}]$ to the SU and to the FEEU temperature interface have been specified. The driving terms are the same as in subsection \ref{sect:scale}. 
In addition to these terms, parasitic forces depending on the temperature must be taken into account.

Residual gas molecules hit the test-mass and generate a radiometer effect force that can be expressed as a disturbing acceleration
\begin{equation}
\frac{1}{2}P_r \frac {S_{\rm TM}}{m_{i} T_{\rm SU}}\alpha(f)  \nabla T_{\rm SU}L_{\rm TM},
\end{equation} 
with the SU internal pressure specified to $P_{\rm{r}}=10^{-5}$\,Pa, with $S_{\rm TM}$ the area along $X$ of the test-mass of mass  $m_i$ and length $L_{\rm TM}$, with $\alpha(f)$ the thermal filtering depending on the frequency of measurement and $\nabla T_{\rm SU}$ the temperature gradient along the $X$-axis. 

Similarly, photons hit the test mass with a radiation pressure which accelerates the test-mass by
\begin{equation}
\frac{16}{3}\sigma_b T_{\rm SU}^3\frac {S_{TM}}{m_{i} c}\alpha(f)  \nabla T_{\rm SU}L_{\rm TM},
\end{equation}
with $\sigma_b$ the Boltzmann constant and $c$ the light velocity.

The thermal filtering is considered between the SU interface (where temperature probes are) and the electrode set surrounding the test-mass from which thermalised molecules or photons leave the surface to hit the test-mass.

The bias can vary because of the Contact Potential Difference sensitivity to temperature which is specified to 15\,$\mu$V\,K$^{-1}$.
It can also vary because of the variation of the electronics voltages (the AC and DC test-mass voltages --$V_p$ and $V_d$--, the actuation voltages, the secondary power voltage). For instance, the requirement on $V_p$ has been set to $8\mu$V\,K$^{-1}$, the  bias stability actuation voltage is specified to better than $2\mu$V\,K$^{-1}$ and the actuation gains specified to $2\times{}10^{-4}$\,K$^{-1}$ for the most constraining specifications. A particular care has also been taken to limit the variations of the power supply at $f_{EP}$ in order to minimise the variation of the power dissipation inside the FEEU \cite{hardycqg6}.

The bias due to the electrostatic stiffness is proportional to \cite{chhuncqg5}
\begin{equation}
\frac {S_{\rm TM}}{e_j m_k}(V_p^2+V_d^2),
\end{equation}
with $e_x$ the gap between the $j$th test-mass and any grounded surface. This term is sensitive to the SU temperature because of the thermal expansion of the electrode set or of the test-mass and sensitive to the FEEU temperature because of the test-mass voltage sensitivity. 

Finally, the bias due to the gold wire stiffness varies with the SU temperature through the thermal expansion of the gold wire and its Young modulus sensitivity to temperature.

The budget of the temperature sensitivity is summarised in Table \ref{tab_B}. 

\begin{table}
\caption{\label{tab_B} Thermal sensitivity requirements of the instrument on $X$-axis.}
\begin{indented}
\item[]\begin{tabular}{@{}lcl}
     \br
     \textbf{Component} & SU & FEEU \cr
       & \multicolumn{2}{c}{Temp. sensitivity} \cr
     \mr
   Scale factor & $8\times{}10^{-6}$\,K$^{-1}$& $2\times{}10^{-6}$\,K$^{-1}$\cr
     \mr
   Alignment & $10^{-7}$\,K$^{-1}$& $2\times{}10^{-8}$\,K$^{-1}$\cr
     \mr
 Lin. Coupling & $5\times{}10^{-8}$\,K$^{-1}$& $6\times{}10^{-9}$\,K$^{-1}$\cr
     \mr
 Ang. Coupling & $5\times{}10^{-9}$\,K$^{-1}$& $6\times{}10^{-10}$\,K$^{-1}$\cr
     \mr
 Bias & $10^{-12}$\,m\,s$^{-2}$K$^{-1}$& $10^{-13}$\,m\,s$^{-2}$K$^{-1}$\cr
    \mr
Bias sensitivity to & & \cr
temperature gradients &\multicolumn{2}{c} {$2\times{}10^{-12}$\,m\,s$^{-2}$(K/m)$^{-1}$}\cr
     \br
\end{tabular}
\end{indented}
\end{table} 

\subsection{Drag-Free} \label{sect:dfacs}
The specification on the linear acceleration level applied to the accelerometer is driven by the DFACS control residuals. The leading component is similar to the instrument parameter $[A^{(d)}]\gamma^{(c)}$. But here we focus on the stochastic and systematic variations of $\gamma^{(c)}$ detailed in Table \ref{Tab_DF}. The scale factor matching and the alignments calibrated to about $10^{-4}$ accuracy allow us to approximate $[A^{d}]$. 

\begin{table}
\caption{\label{Tab_DF} Drag free requirements along $X$, $Y$ and $Z$.}
\begin{indented}
\item[]\begin{tabular}{@{}lcl}
     \br
     \textbf{Frequency} & Specification & Applicability\cr
     \mr
   Random about $f_{\rm{EP}}$ & $3\times{}10^{-10}$\,m\,s$^{-2}$\,Hz$^{-1/2}$ & EP test and calibration\cr
     \mr
   At $f_{\rm{EP}}$, $2f_{\rm{EP}}$ and $3f_{\rm{EP}}$ &$10^{-12}$\,m\,s$^{-2}$ & EP test in inertial pointing \cr
     \mr
   At $f_{\rm{EP}}$ &$10^{-12}$\,m\,s$^{-2}$ & EP test in rotating mode \cr
   At $2f_{\rm{EP}}$ and $3f_{\rm{EP}}$ &$10^{-11}$\,m\,s$^{-2}$ &  \cr
     \mr
   At $f_{\rm{cal}}$, $2f_{\rm{cal}}$ and $3f_{\rm{cal}}$ &$5\times{}10^{-12}$\,m\,s$^{-2}$ & Calibration in inertial pointing \cr
     \br
\end{tabular}
\end{indented}
\end{table}

\subsection{Magnetic field variations} \label{sect:mag}

The analysis of the magnetic effects on the acceleration of the test-mass is detailed in Ref. \cite{hardycqg6}. An allocation of $4\times{}10^{-16}$\,m\,s$^{-2}$ at $f_{\rm EP}$ is considered for the difference of acceleration, and $8\times{}10^{-15}$\,m\,s$^{-2}$ at $2f_{\rm EP}$.

\subsection{Non linearities} \label{sect:nonLin}
On the $X$-axis, a non-linearity can be modelled by an adding the term 
\begin{equation}
4 K_{2x}^{(c)} {\gamma_x^{(c)}}{\gamma_x^{(d)}}+K_{2x}^{(d)} \left(4{\gamma_x^{(c)}}^2+{\gamma_x^{(d)}}^2\right)
\end{equation}
to Eq. (\ref{eq_diffmode}), where $K_{2x}^{(c)}$ and $K_{2x}^{(c)}$ are respectively the common and differential measurement quadratic term. By considering the DFACS performance specification above, the resulting specification on non-linear term has been established with an allocation of 4 times $2\times{}10^{-16}~{\rm m\,s}^{-2}$ at $f_{\rm EP}$ for the two main terms in $\gamma_x^{(c)}$. This leads to specify $K_{2x}^{(c)}$ and $K_{2x}^{(d)}$ to be lower than $14000~{\rm s^2m}^{-1}$.


\section{A posteriori budget from in-flight measurement} \label{sect_com}

Table \ref{tab_budgetV2} has been updated after the commissioning phase. Several inputs to the budget of error had to be modified. First, the instrument bias was greater than expected due to the higher gold wire stiffness. That led to use the DFACS channel for the $X$-axis science measurement instead of the particular science channel. To cope with the necessity of a higher ratio measurement range over bias, the electrode voltages were biased by a constant in order to change the scale factor \cite{liorzoucqg2}. That also implied updating the instrument's control laws. 
Secondly, this stiffness also induces a higher damping of the gold wire and thus a higher noise. By increasing the frequency rate of the satellite rotation (V3 mode), it is possible to take advantage of the decreasing effect of the noise with frequency and of a better temperature filtering.
The thermal sensitivity is also higher than expected but fortunately, the variation of temperature is much lower than considered in Table \ref{tab_budgetV2} because of the higher rotation rate of the satellite: this topic is detailed in Ref. \cite{hardycqg6}. 
Table \ref{tab_budgetV3} takes into account the new flight operation configuration for the satellite and the instrument and their associated environment. The random noise is directly deduced from accelerometer in flight measurement and implicitly includes all error sources. The DFACS performance results from the analysis of the accelerometer and the star sensor outputs combined to a fine trajectography in Ref.  \cite{robertcqg3}. It corresponds to the budget actually estimated in orbit, with the real instrument configuration and noise performances in the case of the SUEP instrument. 

\begin{table}
\caption{\label{tab_budgetV3} Expected budget of performance updated after commissioning phase, with updated data from instrument and DFACS in V3 mode.}
\begin{indented}
\item[]\begin{tabular}{@{}lcc}
     \br
     \textbf{Error source} & \textbf{Contribution in} & \textbf{Contribution in}  \\
      SUEP in mode V3     & \textbf{Random noise} & \textbf{Harmonic error}\\  
                        & m\,s$^{-2}$\,Hz$^{-1/2}$ & m\,s$^{-2}$\\
     \mr
   Earth gravity gradients & & $2.6\times{}10^{-16}$ \\
     \mr
    Instrument gravity &   & $2.0\times{}10^{-16}$ \\
     \mr
    Satellite gravity gradients &  & $1.7\times{}10^{-16}$\\
     \mr
    Angular motions & & $3.8\times{}10^{-16}$\\
     \mr
    Instrument parameter variations &  & $18.6\times{}10^{-16}$\\
     \mr
    Accelerometer measurement noise & $404\times{}10^{-13}$ & \\
    and parasitic forces                      &&\\
     \mr
    Temperature variations &  & $41.2\times{}10^{-16}$\\
     \mr
    Drag-Free Residues &   & $1.7\times{}10^{-16}$\\
     \mr
    Magnetic sensitivity &  & $4.0\times{}10^{-16}$\\
     \mr
    Non linearity & & $3.8\times{}10^{-16}$\\
     \br
   Total & $404\times{}10^{-13}$ & $79.5\times{}10^{-16}$\\
     \br
   \textbf{EP test budgeted for} $\delta : 6.1\times{}10^{-15}$& over 120 orbits& \\
   \textbf{\hspace*{2.3cm} and for} $\delta : 2.0\times{}10^{-15}$& over 1400 orbits& \\
    with $g=7.9$\,m/s$^2$ &&\\

     \br
\end{tabular}
\end{indented}
\end{table}
 
\section{Conclusion} \label{sect_ccl}
The mission MICROSCOPE has been designed to fulfil the objective of a test at $10^{-15}$ accuracy level. All requirements on satellite subsystems have been defined to achieve this objective. As shown, a detailed analysis of all error sources leads to establish a budget of error for the mission. This analysis was updated during the whole mission development duration and served as a guideline to make compromises when it was necessary. This tool was shared within the mission team and in particular within the satellite and instrument team in the framework of the MICROSCOPE performance group which has been meeting monthly for more than 15 years.
The last version of the mission error budget was established with the analysis of the inflight performance. As some subsystem could not be fully tested on ground in their flight configuration, reaching the objective was a very challenging task. Also some unexpected events and breakdowns led the team to reconsider the mission scenario \cite{rodriguescqg4} in order to preserve firstly the integrity of all systems and secondly to get the best possible performance despite difficulties. Ref. \cite{liorzoucqg2, robertcqg3, chhuncqg5, hardycqg6} detail the analysis of the major subsystem errors that finally leads to reach an accuracy test of about $2\times{}10^{-15}$ by considering 1400 orbits of measurement. This figure is not far from the obtained result when each session performance is considered and cumulated \cite{metriscqg9}. It is almost an improvement by a factor 6 compared to the first published results \cite{touboul17, touboul19}.
The long path to this success gives precious lessons that should help to design a new mission with much better accuracy.


\ack

The authors express their gratitude to all the different services involved in the mission partners and in particular CNES, the French space agency in charge of the satellite. This work is based on observations made with the T-SAGE instrument, installed on the CNES-ESA-ONERA-CNRS-OCA-DLR-ZARM MICROSCOPE mission. ONERA authors’ work is financially supported by CNES and ONERA fundings.
Authors from OCA, Observatoire de la C\^ote d’Azur, have been supported by OCA, CNRS, the French National Center for Scientific Research, and CNES. ZARM authors’ work is supported by the DLR, German Space Agency,  with funds of the BMWi (FKZ 50 OY 1305) and by the Deutsche Forschungsgemeinschaft DFG (LA 905/12-1). The authors would like to thank the Physikalisch-Technische Bundesanstalt institute in Braunschweig, Germany, for their contribution to the development of the test-masses with funds of CNES and DLR.

\section*{References}
\bibliographystyle{iopart-num}
\bibliography{biblimscope}

\end{document}